\documentclass[reprint,twocolumn,superscriptaddress,nofootinbib,amsmath,amssymb]{aastex63}

\usepackage{amsmath}
\usepackage{amssymb}
\usepackage{times}
\usepackage{bm}
\usepackage{color}
\newcommand{\ie}{i.e.,~}
\newcommand{\eg}{e.g.,~}

\shorttitle{Microphysical plasma relations from special-relativistic
  turbulence}
\shortauthors{Meringolo et al.}
\graphicspath{{./}{figures/}}

\begin{document}

\title{Microphysical plasma relations from kinetic modelling of
  special-relativistic turbulence}

\correspondingauthor{Claudio Meringolo, Alejandro Cruz-Osorio}
\email{claudiomeringolo@unical.it, osorio@itp.uni-frankfurt.de}

\author[0000-0001-8694-3058]{Claudio Meringolo}
\affiliation{Institut f\"ur Theoretische Physik, Goethe Universit\"at, Frankfurt, Germany}
\affiliation{Dipartimento di Fisica, Universit\`a della Calabria, I-87036 Cosenza, Italy}

\author[0000-0002-3945-6342]{Alejandro Cruz-Osorio}
\affiliation{Institut f\"ur Theoretische Physik, Goethe Universit\"at, Frankfurt, Germany}

\author[0000-0002-1330-7103]{Luciano Rezzolla}
\affiliation{Institut f\"ur Theoretische Physik, Goethe Universit\"at, Frankfurt, Germany}
\affiliation{Frankfurt Institute for Advanced Studies, Frankfurt, Germany}
\affiliation{School of Mathematics, Trinity College, Dublin, Ireland}

\author[0000-0001-8184-2151]{Sergio Servidio}
\affiliation{Dipartimento di Fisica, Universit\`a della Calabria, I-87036 Cosenza, Italy}



\begin{abstract} 
  The microphysical, kinetic properties of astrophysical plasmas near
  accreting compact objects are still poorly understood. For instance, in
  modern general-relativistic magnetohydrodynamic simulations, the
  relation between the temperature of electrons $T_{e}$ and protons
  $T_{p}$ is prescribed in terms of simplified phenomenological models
  where the electron temperature is related to the proton temperature in
  terms of the ratio between the gas and magnetic pressures, or $\beta$
  parameter. We here present a very comprehensive campaign of
  {two-dimensional} kinetic Particle-In-Cell (PIC) simulations of
  special-relativistic turbulence to investigate systematically the
  microphysical properties of the plasma in the trans-relativistic
  regime. Using a realistic mass ratio between electrons and protons, we
  analyze how the index of the electron energy distributions $\kappa$,
  the efficiency of nonthermal particle production $\mathcal{E}$, and the
  temperature ratio $\mathcal{T}:=T_{e}/T_{p}$, vary over a wide range of
  values of $\beta$ and $\sigma$. For each of these quantities, we
  provide two-dimensional fitting functions that describe their behaviour
  in the relevant space of parameters, thus connecting the microphysical
  properties of the plasma, $\kappa$, $\mathcal{E}$, and $\mathcal{T}$,
  with the macrophysical ones $\beta$ and $\sigma$. In this way, our
  results can find application in wide range of astrophysical scenarios,
  including the accretion and the jet emission onto supermassive black
  holes, such as M87* and Sgr A*.
\end{abstract}



\section{Introduction}
\label{sec:intro}

Considerable effort has been dedicated over the last few years to the
modeling via general-relativistic simulations of plasma accreting onto
supermassive black holes \citep{nathanail2020, DelZanna2020,
  Ripperda2019, Younsi2020, Dihingia2022} and neutron stars
\citep{Parfrey2017, Abarca2018, Das2022, Cikintoglu2022}. Among the
different approaches considered, surely general-relativistic
magnetohydrodynamics (GRMHD) simulations have been the focus of many
groups worldwide~\citep[see, \eg][]{EHT_M87_PaperV, Porth2019,
  EHT_SgrA_PaperV}. While essential to make theoretical progress on these
scenarios, these GRMHD simulations can only describe the dynamically
important part of the fluid, the protons (or ``ions'' as they are
sometimes referred to), leaving completely undetermined the physical
properties -- such as the energy distribution, the number densities, and
the temperatures -- of the ``lighter'' part of the fluid, namely, the
electrons. This represents a serious limitations for two different
reasons. First, in hot, ionized plasma jets around black holes, the
Coulomb coupling between electrons and protons is inefficient, {so
  that protons and electrons are likely to have distinct temperatures, as
  it happens in the solar wind~\citep{Tu1997, vanderHolst2010,
    Howes2010, Dihingia2022}}.  Second, a proper knowledge of the
electron energy distribution is essential in order to obtain accurate
imaging of supermassive black holes and hence compare with the
observations~\citep{Davelaar2019, Mizuno2021, Cruz2022}.

To cope with this problem, a number of phenomenological prescriptions
have been suggested in the literature to relate the electron temperature
to the simulated proton temperature. In this context, a very commonly
employed approach is the so-called $R\!-\!\beta$ model
\citep{Moscibrodzka2016}, where the electrons temperature is related to
the protons temperature in terms of the plasma-$\beta$ parameter, \ie the
ratio of the thermal-to-magnetic pressure, and of two free parameters,
$R_{\rm low}$ and $R_{\rm high}$~\citep[see also][for a critical-$\beta$
  model, where two additional parameters are
  introduced]{Anantua2020}. The $R\!-\!\beta$ approach has been widely
used by the Event Horizon Telescope (EHT) Collaboration to reconstruct
theoretically the first images of supermassive black holes
M87*~\citep{EHT_M87_PaperI}, and Sgr A*~\citep{EHT_SgrA_PaperI}. These
investigations, in particular, have resorted to a simplified version of
the $R\!-\!\beta$ approach in which $R_{\rm low}=1$ and spanning
different values of $R_{\rm high}$~\citep{EHT_M87_PaperV,
  EHT_SgrA_PaperV}. Taking into account a more realistic description of
the plasma parameters using self-consistent kinetic models has shown that
finer details of the image can appear, but also that the $R\!-\!\beta$
approach is remarkably robust~\citep{Mizuno2021}.

{Clearly, it is essential to connect the microphysical properties of
  the plasma with the macrophysical ones $\beta$ and $\sigma$, where
  hybrid-kinetic models might have some limitations
  \citep{Arzamasskiy2019, Valentini2014, Cerri2017}.} To this scope, we
have performed $38$ large-scale fully kinetic (\ie both protons and
electrons are treated as particles) Particle-In-Cell (PIC) simulations of
special-relativistic plasma in the so-called ``trans-relativistic regime'',
that is, when the plasma magnetization $\sigma$ -- the ratio between the
magnetic energy density to enthaply density (see below for a definition) 
-- is of order
unity ~\citep{Ripperda2019, Mizuno2021, Bandyopadhyay2022, Janssen2021},
and covering four orders of magnitude in the plasma-$\beta$ parameter
{(see Appendix for details on the various simulations)}. In all
simulations, we employ a physical proton-to-electron mass ratio
\citep[see][for the importance of using a realistic mass
  ratio]{Rowan2017}, and analyze the most important microphysical
properties of the turbulent plasma, namely, the spectral index of the
electron energy distributions $\kappa$, the efficiency in the production
of nonthermal particles $\mathcal{E}$, and the temperature ratio
$\mathcal{T}:=T_{e}/T_{p}$.  {The parameter ranges explored here
  overlap and extend those considered in previous and influential works
  of astrophysical kinetic turbulence \citep{Zhdankin2019, Zhdankin2021,
    Kawazura2019, Kawazura2020}.}

Exploiting the large coverage of the space of parameters, we are able to
model {via analytic fitting functions} the behaviour of all of these
quantities, thus providing a convenient tool to introduce kinetic effects
in global GRMHD simulations of accretion onto compact objects, thus
improving the modeling of radiatively inefficient accretion flows around
black holes, such as M87* or Sgr A*~\citep{Tchekhovskoy2012, Qian2018,
  Porth2019, Cruz2022, Chatterjee2021, Ripperda2022, EHT_SgrA_PaperV}.

\section{Simulation setup}

To carry out our investigation and to simulate the full development of
relativistic turbulence in kinetic plasmas, we use the publicly available
\texttt{Zeltron} code \citep{Cerutti2015, Cerutti2019}. In particular, we
employ a two-dimensional (2D) geometry in Cartesian coordinates
retaining, as in any fully consistent plasma model, the three-dimensional
components of the magnetic and electric fields, of the current density,
and of the pressure tensor. The temperature of each species, $\alpha={e,
  p}$ for electrons and protons, is specified through the plasma-$\beta$
parameter $\beta_\alpha := 8 \pi n_\alpha k_B T_\alpha / B_0^2$, where
$n_\alpha$ and $T_\alpha$ are the number densities and the temperatures,
respectively. Here, ${\bm B}_0=(0,0,B_0)$ is the magnetic-field vector in
the ambient plasma ($B_0={\rm const.}$), and $k_B$ is the Boltzmann
constant.

We initialize all of our simulations with the same number density for
electrons and protons in a computational domain that is a Cartesian box
of side $L_x = L_y = L = 16384 \,dx$, where $dx = d_e/3$ is the cell
resolution and $d_e := c/ \omega_{pe}$ is the electron-skin depth. In the
above, $c$ is the speed of light and {$ \omega_{pe} := \sqrt{(4 \pi
    n_e e^2)/m_e} \left[ 1 + {\theta_e}/({\Gamma_e -1}) \right]^{-1/2} $
  is the electron plasma frequency, $m_e$ the electron mass, $\Gamma_e$
  the electron adiabatic index, and $\theta_e := k_B T_e/m_e c^2$ the
  dimensionless electron temperature.} We have also carried out three
more simulations with a smaller box, $L=2730\, d_e$, and resolution of
$(8192)^2, (16384)^2$, and $(32768)^2$ points {(see Appendix for
  details)}. Furthermore, we set up our computational box so that it is
periodic in the $x$- and in the $y$-directions. Finally, in all our
simulations, each computational cell is initialized with $10$ particles
per cell (\ie 5 electrons and 5 protons). As a result, during our
evolutions we follow the dynamics of $\sim 2.7 \times 10^9$ particles per
simulation.

We initialize our system as done in typical simulations of plasma
turbulence~\citep[see, \eg][]{Servidio2012}. The initial conditions
consist of a relativistic plasma perturbed by a 2D spectrum of Fourier
modes for the magnetic field. To avoid any compressive activity, neither
density perturbations nor parallel variances (\ie with components out of
plane) are imposed at $t=0$. In practice, we start from expressing the
$z$-component of the vector potential in Fourier modes as $A_z(x,y) :=
\sum_{k_x, k_y} A_k \exp{[i(\bm{k} \cdot \bm{x} +\phi_{\bm{k}})]}$, where
$\bm{k}=(k_x, k_y)$ is the wavevector with modulus $k=|\bm{k}| =(2\pi/L)
m$ ($m$ is the dimensionless wavenumber), and $ \phi_{\bm{k}}$ are
randomly chosen phases. The amplitude of the modes is set as $A_k=
\left[1+(k/k_0)^{15/3} \right]^{-1}$, such that it is peaked at $k_0 =
(2\pi/L) m_0$ with $m_0 = 4$. The spectrum is set to zero at $m>7$ in
order to construct initial conditions consistent with random large-scale
structures. The magnetic-field components $B_x$ and $B_y$ are then
computed by straightforward derivatives. Finally, to explore a 
regime of strongly perturbed field lines, we fix the 
amplitude of the fluctuations to be $\langle B_\perp \rangle /B_0 
\sim 1$, where $\langle B_\perp \rangle$ is the root-mean-square 
value of the in-plane fluctuations. This choice leads to a broader 
particle energy distribution, while in \cite{Nattila2022} the authors 
showed that when the amplitude is small, the particle energy distribution 
is quasi-thermal.

\begin{figure}
\includegraphics[height=74mm,width=86mm]{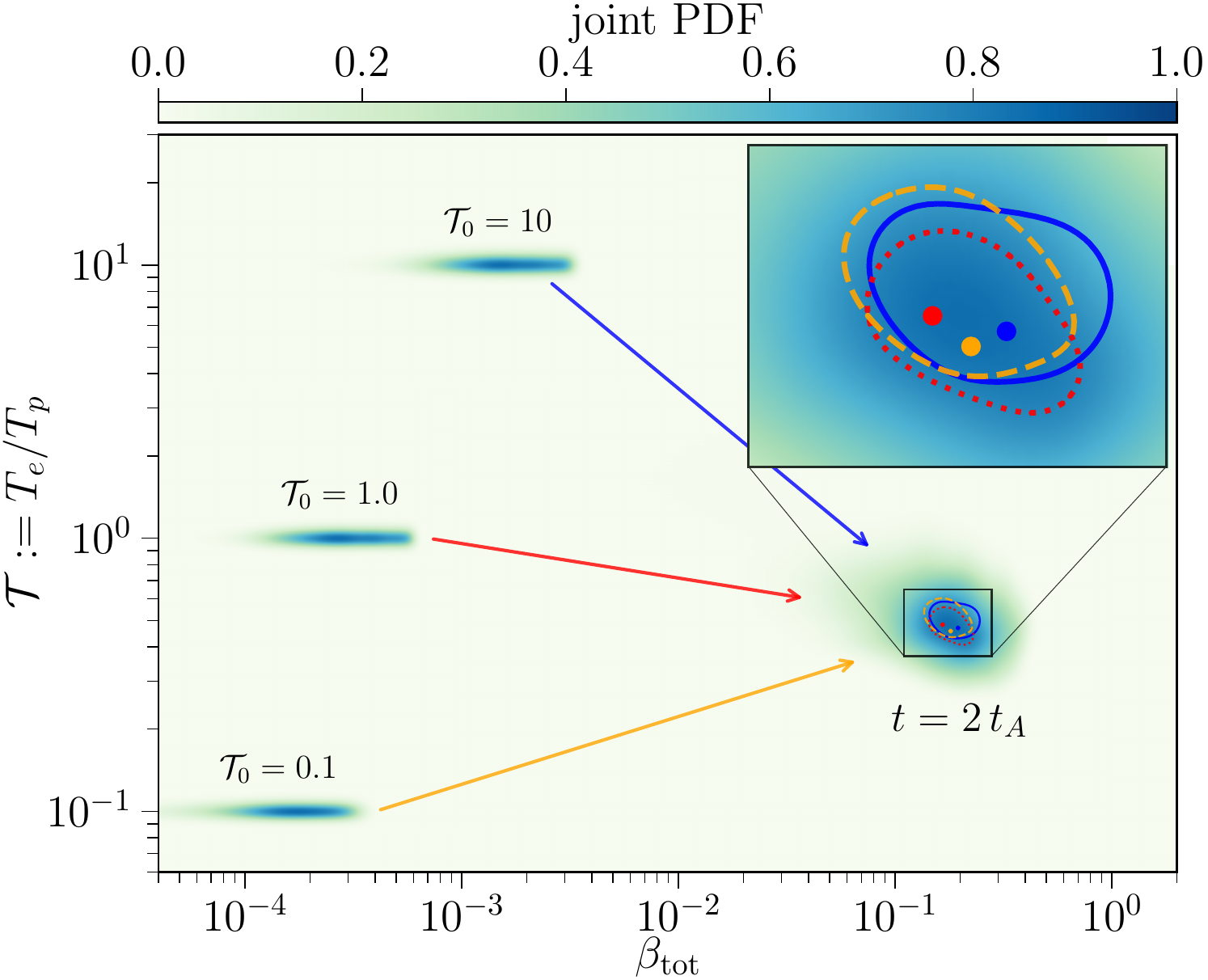}
\caption{Initial and final values at time $t=2\,t_A$ of the normalized
  joint PDFs of the temperature ratio $\mathcal{T}$ and of the total
  $\beta$ parameter, $\beta_{\rm tot} := \beta_{e} + \beta_{p}$.  The
  data refers to three representative simulations with initial
  temperature ratio $\mathcal{T}_0=0.1,1.0$ and $10$. The inset shows the
  $90\%$ contour lines of the joint PDFs, while the circles mark the
  maxima of each PDF. Note that all PDFs converge to the same final area
  in the $(\mathcal{T},\beta_{\rm tot})$ plane despite the very different
  initial data.}
\label{fig:1} 
\end{figure}

Other quantities that will be referred to in the rest of the paper are
the Alfv\'en crossing time as $t_{A} := L/v_A$, where
$v_A:=c\sqrt{\sigma/(1+\sigma)}$ is the Alfv\'en speed. The plasma
magnetization is instead defined as $\sigma:=B_0^2/(4 \pi w)$, where $w$
is the enthalpy density of the plasma $w:=(\rho_e + \rho_p)c^2 + \Gamma_e
\epsilon_e + \Gamma_p \epsilon_p$, with $\rho_{e,p}$ and $\epsilon_{e,p}$
being, respectively, the rest-mass densities and the internal energy
densities of electrons and protons when following an ideal-fluid equation
of state~\citep{Rezzolla_book:2013}.

As the simulation proceeds, turbulent magnetic reconnection takes place,
leading to a nonlinear change in magnetic topology and converting
magnetic energy into kinetic and internal energy. This process strongly
affects the dynamics of the plasma on all the scales we could reproduce
with our simulations. This highly dynamical system evolves with magnetic
flux ropes moving, colliding, and sometimes repelling each other
depending on the magnetic-field polarity. This dynamics proceeds till a
stationary state is achieved after about an Alfv\'en crossing time (see
also the top panels of Fig.~\ref{fig:2} {and the Appendix for a
  detailed discussion}).

\section{Results}
\label{sec:results}

At the initial time, after fixing $\beta$ and $\sigma$ for each
simulation, we set the temperatures, which are uniform for both
species. In particular, we first specify the proton-$\beta_p$ parameter
and then obtain the electron temperature so as to have a specific initial
temperature ratio $\mathcal{T}_0 := T_{e,0}/T_{p,0}$. At any time during
the simulation we measure the spatial distributions in the $(x,y)$ plane
of $\beta_{\text{tot}} := \beta_e + \beta_p$ and $\mathcal{T}$, from
which we compute the joint PDFs for the two quantities, namely,
$(\beta_{\text{tot}}, \mathcal{T})$.
{The temperature for each species is computed from $ T_{\alpha} :=
p_{\alpha} / n_{\alpha} k_B $, where $p_{\alpha}$ is the isotropic
pressure, \ie $p_{\alpha} := \tfrac{1}{3} (p_{\alpha}^{xx} +
p_{\alpha}^{yy} + p_{\alpha}^{zz})$ and $p_{\alpha}^{ij}$ is the pressure
tensor}.

This is shown in Fig.~\ref{fig:1}, where we report the joint PDFs at the
initial ($t=0$) and final ($t=2\,t_A$) times for three representative
simulations initialized respectively with $\mathcal{T}_0 = 0.1, 1.0$, and
$10$. Clearly, the three initial setups have different joint PDFs
narrowly distributed around the three initial values of the temperature
ratio $\mathcal{T}_0$. Interestingly, however, at the final time they
have all converged to the same equilibrium distribution, irrespective of
the initial data. This can be best appreciated in the inset, which
reports a zoom-in of the central region of the final distributions, with
the color-coded contour reporting the $90\%$-value for each simulation,
while the circles represent the maximum of each joint PDF. This
convergence has been verified to take place for four different values of
the initial temperature ratio ($\mathcal{T}_0=0.01, 0.1, 1.0$ and
$10.0$), while keeping $\sigma=0.3$ and $\beta=3 \times 10^{-4}$. The
behaviour in Fig.~\ref{fig:1} {induces us to conjecture} that the choice
of the initial temperature $\mathcal{T}_0$ is effectively unimportant at
least in the ranges explored here\footnote{{A word of caution: we have
  shown the initial temperature to be irrelevant once turbulence is
  developed for a specific set of initial values of $\beta$ and
  $\sigma$. Given the physical arguments given above, extending this
  conclusion to different initial values is a conjecture that is
  reasonable but challenging to prove, especially for $\beta \sim 1$.}}
as its memory is lost by the time the system has reached a steady
state. In view of this, we set $\mathcal{T}_0 =1.0$ for the 35
simulations performed varying $\sigma$ and $\beta$ (note that with such
initial temperature ratio, the plasma-$\beta$ parameter is the same for
electrons and protons, \ie $\beta_e = \beta_p =: \beta$).  {The ranges of
  $\sigma$ and $\beta$ explored are compatible with previous kinetic
  studies, state-of-the-art GRMHD simulations, and radiative-transfer
  calculations \citep{Ball2018a, Cruz2022, Fromm2021b}.} As also noted by
\citet{Pecora2019}, higher values of $\beta$ would require a much higher
number of particles to counter the statistical noise, making purely
  PIC calculations of this type computationally expensive with modern
resources.

\begin{figure*}
\includegraphics[height=128mm,width=178mm]{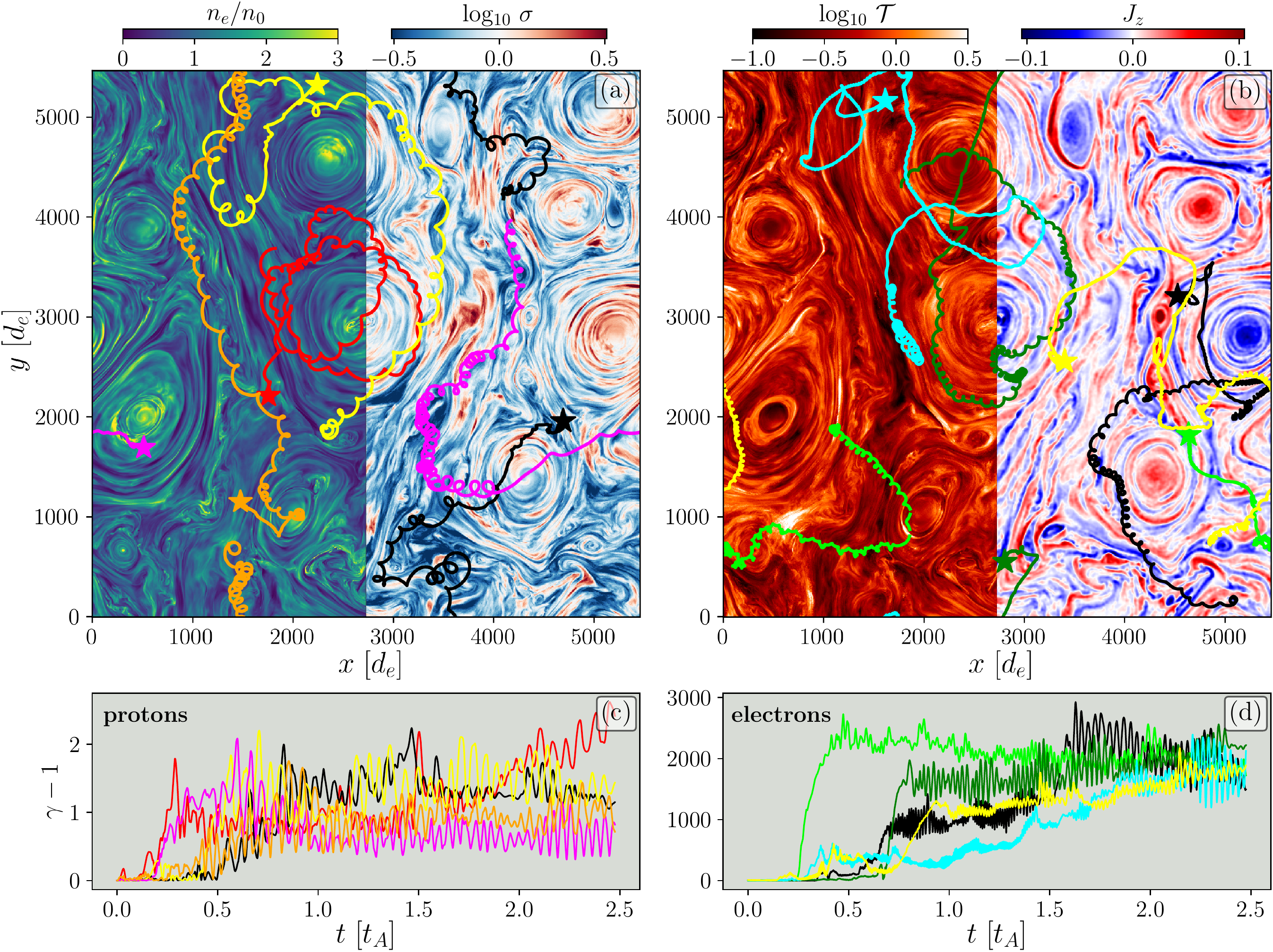}
\caption{Representative quantities in a fully developed 2D turbulence at
  $t = 1.5\,t_A$ for a representative simulation with $\sigma=1$ and
  $\beta=3 \times 10^{-3}$. The top panels offer a dual view of: the
  electron number density normalized to the initial value $n_e/n_0$ and
  of the magnetization $\sigma$ [panel (a)], and of temperature ratio
  $\mathcal{T}$ and of the total current density $J_z$ [panel (b)]. Also
  overplotted with different colors are representative particle
  trajectories, with protons on the left and electrons on the right of
  each panel (the initial position of each particle is marked with a
  star). The lower panels [(c) and (d)] report instead the evolution of
  the Lorentz factor for the same particles marked above.}
\label{fig:2}
\end{figure*}

Figure~\ref{fig:2} provides a very compact but powerful overview of the
fully developed turbulent state for a simulation with $\sigma=1.0$ and
$\beta=3 \times 10^{-3}$, at time $t = 1.5\,t_A$. Each upper panel is
split into two regions reporting different plasma properties. Panel (a)
shows the electron number density $n_e$ normalized to the initial number
of particles per cell $n_0$ (left), and the magnetization $\sigma$
(right). Panel (b), instead, reports temperature ratio $\mathcal{T}$
(left) and the out-of-plane electric-current density $J_z$ (right).  
Note how, in analogy to nonrelativistic kinetic simulations,
vortex-like and sheet-like structures corresponding to magnetic flux
tubes are present at all the scales that are resolved in the
simulation~\citep{Servidio2012, Comisso2018, Parashar2018,
  Pecora2019}. High number-densities ``magnetic islands'' can be found in
large-scale flux tubes, and in general, the density is larger in these
coherent quasi-circular structures.

At the same time, the largest temperatures (ratios) are not achieved at
the center of the islands, which are instead comparatively cooler.  This
is because the temperature is higher between flux tubes, where
reconnection layers lead to the formation of plasmoids within narrow
current sheets~\citep{Servidio2009, Comisso2018, Pezzi2021}. Elongated
unstable current sheets tend to fragment into chains of plasmoids and
small-size current sheets are appear on a wide range of
scales~\citep{Hellinger2015, Dong2018, Huang2016}. Notice also that the
out-of-plane electric-current density $J_z$ shows a variety of current
sheets of different sizes. Some of these current layers break into
smaller plasmoids and these regions are important for the heating of the
plasma and the acceleration of the particles.

The various quantities shown in Fig.~\ref{fig:2} are overlaid with the
trajectories of some of the most energized particles that we tracked
(protons in the left panels and electrons in the right ones). In
particular, we track a sample of $500$ electrons and $500$ protons during
the whole simulation, both randomly chosen. The starting-position of each
particle is marked with a star. Note how, quite generically, and in
addition to the basic gyrations at the corresponding Larmor radii, there
are particles that have closed orbits as they are trapped in a flux rope,
while others experience turnovers that suddenly bend the trajectory,
similarly to what observed in nonrelativistic turbulence simulations
\citep{Pecora2018}\footnote{{When the turbulence is fully developed,
  the velocity distribution of the electrons is highly nonthermal and
  their Larmor radius is significantly larger as a result of the large
  accelerations, and this effectively
  increases our resolution.}}. Overall, when a particle experiences a
reconnection process and is accelerated, it increases abruptly its Larmor
radius, but also its Lorentz factor $\gamma$, and kinetic energy.

In the lower panels (c) and (d) of Fig.~\ref{fig:2} we show the evolution
of the Lorentz factor of the particles tracked in the upper panels (a)
and (b), with protons being reported in panel (c) and electrons in panel
(d). As expected, and shown by the different vertical scales of panels
(c) and (d), electrons experience considerably larger accelerations when
compared to protons. This is simply due to the different masses of the
two species: electrons, which have smaller Larmor radius, are more
efficiently accelerated by the thin current sheets where magnetic
reconnection takes place. This stochastic acceleration mechanism of
multi-reconnection events is very efficient and commonly observed in
astrophysical plasma turbulence~\citep{Drake2009,
  Haynes2014}.

The tracked particles start from $\gamma \gtrsim 1$, and most of them
experience a sudden acceleration episode, and then a sequence of
second-order Fermi-like processes of acceleration 
{\citep{Comisso2018,Comisso2019}}. Particles trapped in magnetic
islands show a Lorentz factor increasing in time (\eg the red proton in
the left panels). Other particles, instead gain energy only once and then
reach a quasi-steady state as is typical for particles entering the
magnetic island only for a short time and then being bounced in a
stochastic manner between different structures.

\begin{figure}
\includegraphics[height=110mm,width=86mm]{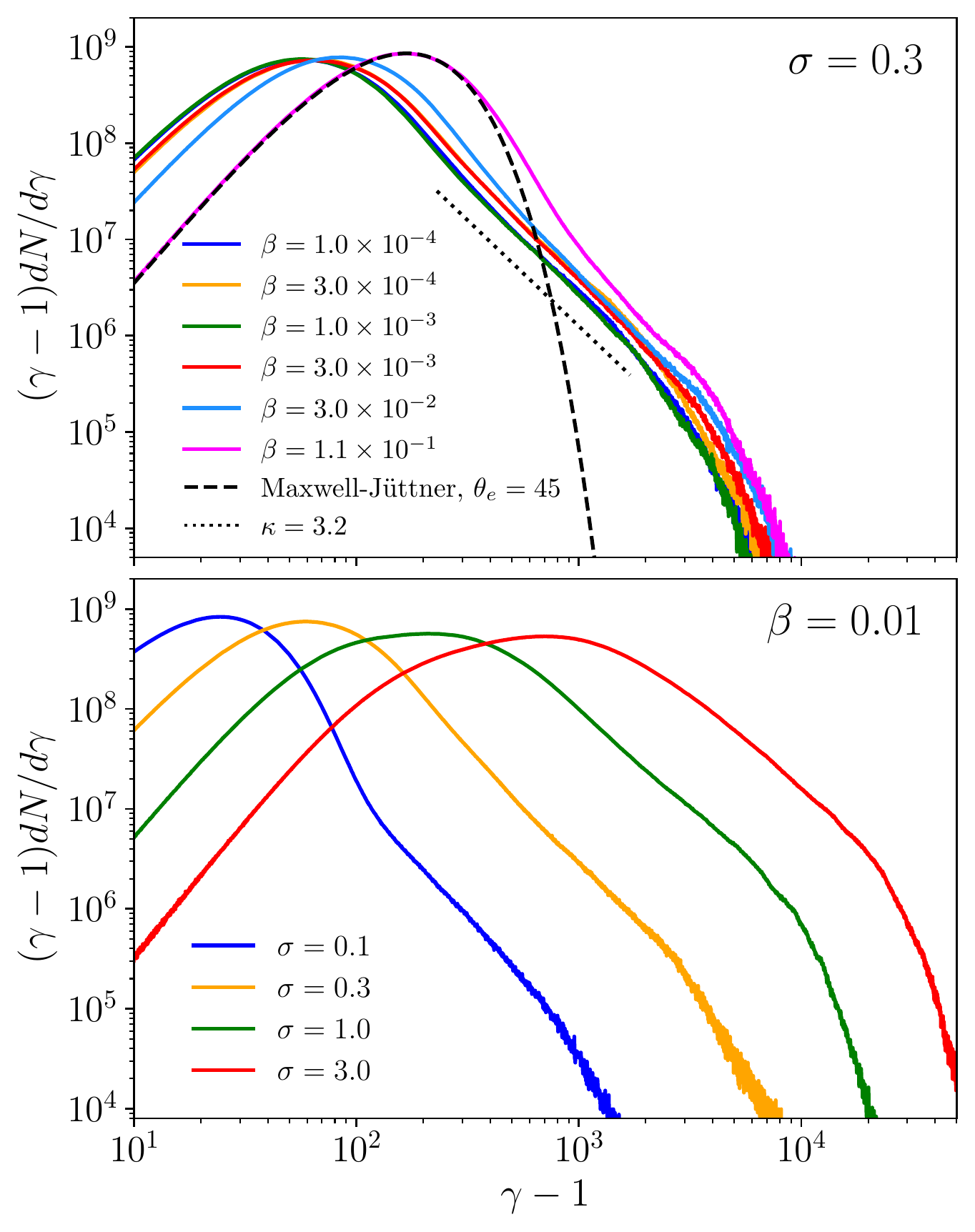}
\caption{\textit{Top panel:} electron-energy spectra at $t=2\,t_A$ for
  simulations with $\sigma=0.3$ and different values of $\beta$;
  indicated with a dashed line is the a Maxwell-J\"uttner distribution
  for $\beta \simeq 0.1$, while the dotted line indicates the almost
  constant spectral index $\kappa\simeq3.2$. \textit{Bottom panel:} Same
  as above, but for simulations with $\beta=0.01$ and different values of
  $\sigma$.
\label{fig:3}}
\end{figure}

\begin{figure*}
\includegraphics[width=1.0\textwidth]{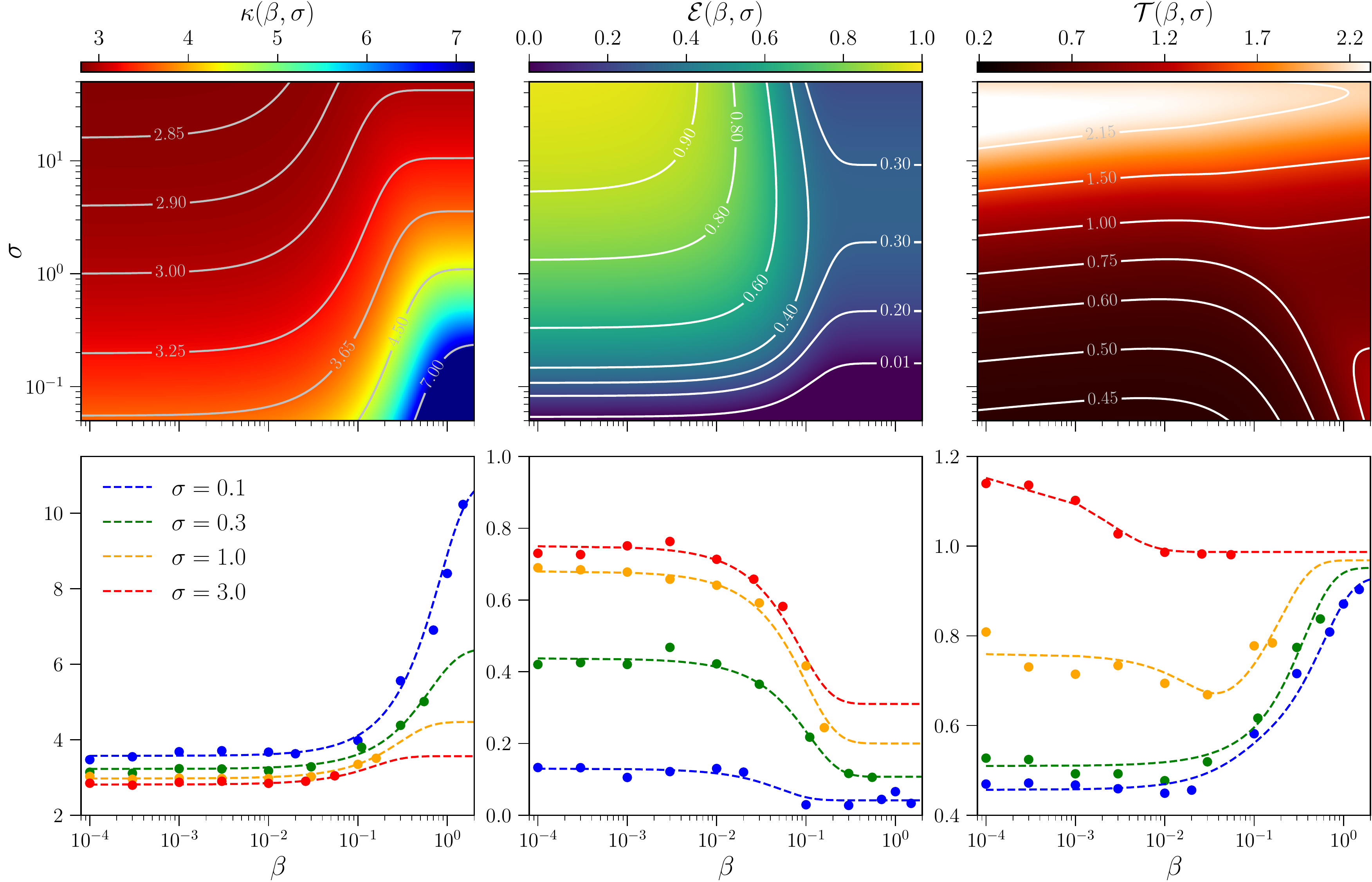}
\caption{\textit{Top panels:} from the left to right are reported as a
  function of $\beta$ and $\sigma$: the electron spectral index $\kappa$,
  the energy efficiency $\mathcal{E}$, and the temperature ratio
  $\mathcal{T}$, respectively [see
    Eqs.(\ref{kappa2D_2})--(\ref{T2D})]. \textit{Bottom panels:} Same as
  above, but at fixed values of the magnetization ($\sigma=0.1-3.0$);
  each circle refers to a distinct simulation.}
\label{fig:4}
\end{figure*}

Relativistic hydrodynamical turbulence naturally provides a landscape of
intermittency and large spatial variance because the compressibility is
enhanced by relativistic effects~\citep{Radice2012b}; in addition,
relativistic magnetohydrodynamical turbulence provides the natural
conditions to produce extreme-acceleration events and to generate a large
population of particles -- electrons in particular -- with energy
distributions that differ significantly from a thermal one~ {
  \citep[see, \eg][]{Zhdankin2017}}. This is summarized in
Fig.~\ref{fig:3}, which reports the electron energy-distribution
functions (spectra) $(\gamma-1) dN/d \gamma$ at $t=2\,t_A$ as a function
of the Lorentz factor $\gamma-1$, for some representative
simulations. More specifically, the upper panel shows the electron
spectra from simulations with $\sigma=0.3$ and for a wide range of values
of $\beta$; the black dashed line is a Maxwell-J\"uttner distribution
where the value of the dimensionless electron temperature $\theta_e:=k_B
T_e/(m_ec^2)=45$ is chosen to reproduce the low-energy part of the spectrum
for the case $\beta=0.11$ and is obviously different for each
simulation. Note that the high-energy part of the spectra is well
approximated by a power-law $dN/d\gamma \propto \gamma^{-\kappa +
  1}$~\citep{Davelaar2019, Fromm2022}, whose index $\kappa\simeq 3.2$ is
quite insensitive to the value of the plasma-$\beta$ parameter in the
range $\beta \lesssim 3 \times10^{-3}$ (see black dotted line). For very
large values of $\beta$, however, a single power law does not represent
the distribution accurately, and only the very high-energy part of the
spectrum maintains an index $\kappa\simeq 3.8$.

In the bottom panel of Fig.~\ref{fig:3}, we instead explore how the
electron-energy spectra change when varying $\sigma$ while keeping
$\beta=0.01$. Note that as the magnetization increases, the amount of
magnetic energy available for dissipation increases, leading to a
systematic shift towards progressively larger energies of the
spectra. Furthermore, the high-energy part of the spectra are well
approximated by power laws with indexes $\kappa\simeq 3-4$, while the
highest regions of the spectra terminate with increasingly harder
slopes. Overall, and in agreement with several previous works
\citep{Comisso2018} -- some of which even have different initial
conditions \citep{Werner2018, Ball2018a} -- our results clearly indicate
that turbulence {promotes} the particle acceleration, producing
energy distributions that contain a considerable fraction of very
energetic (suprathermal) particles.

Given the kinetic behaviour of the plasmas described so far, it is
essential to be able to express their properties {via analytic
  fitting functions} and in terms of the basic parameters of the plasma,
namely, $\beta$, $\sigma$, so that the resulting expressions can then be
employed directly in the GRMHD modelling of astrophysical plasma. A
summary of this analytical modelling is presented in Fig.~\ref{fig:4},
where in the top row we show as a function of $\beta$ and $\sigma$,
respectively, the electron spectral index $\kappa$, the nonthermal energy
efficiency $\mathcal{E}$, and the temperature ratio $\mathcal{T}$. Note
that the data reported in the first two columns refers to simulations at
$t=2\,t_A$, while that in the right column is averaged over the time
window $1.7 < t/t_A < 2.3$ to avoid the oscillations introduced by the
stochastic behavior of turbulence. Similarly, the bottom row of
Fig.~\ref{fig:4} reports one-dimensional cuts of the same quantities, but
at fixed values of the magnetization ($\sigma=0.1-3.0$), where each
circle refers to a distinct simulation of our set.  {Note that for
  any fixed value of $\sigma$ we explored plasma parameters up to the
  maximum one $\beta_{\text{max}} \sim 1/(4\sigma)$ \citep{Ball2018a},
  where our estimates are inevitably less accurate.}

Exploiting the large set of simulations performed, we can now construct
analytical 2D fits to the various quantities, starting with the electron
spectral index $\kappa(\beta,\sigma)$, which can be expressed as

\begin{equation}
   \kappa(\beta,\sigma) = k_0 + \frac{k_1}{\sqrt{\sigma}} + k_2 \,
   \sigma^{-6/10} \,\text{tanh} \left[ k_3 \, \beta \, \sigma^{1/3} \right]
   \,,
   \label{kappa2D_2}
\end{equation}

where $k_0 = 2.8,\, k_1 = 0.2,\, k_2 = 1.6$ and $k_3 = 2.25$ (see
top-left panel of Fig.~\ref{fig:4}). {Note that~\citet{Zhdankin2017}
  have proposed a similar but simpler fitting expression which depends
  $\sigma$ only and thus does not account for variations in the plasma
  $\beta$.} Overall, the spectral index shows two main features. First,
at fixed $\sigma$, the spectral index is essentially independent of
$\beta$, for $\beta \lesssim 10^{-2}$, but it increases at larger values
of $\beta$, approaching a very steep tail. Second, at fixed $\beta$, the
index becomes generally smaller for increasing values of $\sigma$.

Next, we quantify the efficiency in the production of particles with
nonthermal energies in terms of the weighted average of the excess over a
Maxwell-J\"uttner distribution~\citep{Ball2018a}, namely

\begin{equation}
  \mathcal{E} := \frac{\int_{\gamma_{0}}^{\infty} \left[{dN}/{d \gamma} -
      f_{_{\rm MJ}}(\gamma, \theta) \right](\gamma-1)\, d \gamma}{
    \int_{\gamma_{0}}^{\infty} ({dN}/{d \gamma})(\gamma-1)\, d \gamma}\,,
  \label{epsilon}
\end{equation}

where $\gamma_{0}$ denotes the peak of the spectrum, ${f_{_{\rm MJ}}
  := \gamma^2 v/[c\, \theta_e K_2(1/ \theta_e) ] e^{-\gamma /
    \theta_e}}$, with $v$ the velocity and $K_2$ the modified Bessel
function of the second kind. The corresponding 2D fit of the data can
then be expressed as

\begin{equation}
   \mathcal{E}(\beta,\sigma) = e_0 + \frac{e_1}{\sqrt{\sigma}} + e_2 \,
   \sigma^{1/10} \,\text{tanh} \left[ e_3 \, \beta \, \sigma^{1/10}
     \right]\,,
   \label{E2D}
\end{equation}

where $e_0 = 1.0,\, e_1 = -0.23,\, e_2 = 0.5$ and $e_3 = -10.18$ (see
top-middle panel of Fig.~\ref{fig:4}). Also in this case, the energy
efficiency shows three main features. First, for $\beta \lesssim 10^{-2}$
the efficiency saturates at a value that is independent of $\beta$, but
systematically larger for higher values of $\sigma$. Second, for high
values of $\beta$ and low values of $\sigma$, it approaches $\mathcal{E}
\sim 0$, because the electron spectrum becomes significantly
softer. Third, for higher values of $\sigma$, the efficiency is the
largest, since the spectra widen to larger electron energies.
Interestingly, these results are similar to the ones found by
  \cite{Ball2018a} when using different initial conditions.

Finally, we consider what is arguably the most important quantity
modelled in our simulations, namely, the dependence of the temperature
ratio on the plasma properties. The corresponding 2D fit is given by

\begin{eqnarray}
  \mathcal{T}(\beta, \sigma) & = & t_0 + t_1\, \sigma^{\tau_1}
  \tanh\left[t_2 \, \beta \, \sigma^{\tau_2}\right] \nonumber \\ &&
  \phantom{t_0 } + t_2\, \sigma^{\tau_3} \tanh\left[t_3 \,
    \beta^{\tau_4}\, \sigma \right]\,,
  \label{T2D}
\end{eqnarray}

where $t_0 = 0.4,\, t_1 = 0.25,\, t_2 = 5.75,\, t_3 = 0.037$, and
$\tau_1=-0.5,\, \tau_2=0.95,\, \tau_3=-0.3,\, \tau_4=-0.05$ (see
top-right panel of Fig.~\ref{fig:4}). Overall, it is easy to see that for
low magnetizations, \ie $\sigma \in [0.1, 0.3]$, and small values of the
$\beta$ parameter, \ie $\beta \lesssim 0.01$, the temperature ratio is
essentially constant and then starts to grow to values as large as
$\mathcal{T}\simeq1$ for $\beta \lesssim 1.0$. On the other hand, for
high values of the magnetization, \ie $\sigma\simeq3.0$, the behavior is
quite the opposite, the values of $\mathcal{T}$ are higher for lower
$\beta$ and decrease when increasing $\beta$. For intermediate values of
the magnetization, \ie $\sigma=1.0$, the behavior is a combination of the
two described above, showing a nonmonotonic dependence for $\beta \in
[0.01, 0.1]$. Interestingly, in all cases, $\mathcal{T} \sim 1.0$ for
$\beta \simeq 1$, independently of the value of $\sigma$, thus
highlighting that, under these conditions, electrons and protons are
fully coupled and have roughly the same temperature. Conversely, for
$\beta\lesssim10^{-4}$, the temperature ratio will depend on the plasma
magnetization, being larger for larger magnetizations, as expected for
regimes where electrons can be accelerated to suprathermal energies at
reconnection sites. More importantly, expression~\eqref{T2D} provides a
compact and microphysically consistent description of the electron
temperatures that can be employed in modern GRMHD codes of accretion
flows onto black holes.

We conclude the discussion of our results by returning to the behaviour
of the electron spectral index $\kappa$. As shown in the top-left panel
of Fig.~\ref{fig:4} and summarized in Eq.~\eqref{kappa2D_2}, electron
acceleration is higher in low-$\beta$ and high-$\sigma$ turbulent
plasmas. As suggested already by~\citet{Drake2009}, this behaviour may be
due to the interaction of the electron orbits with small-sized current
sheets; such a mechanism can then extract particles from the thermal
population and bring them to very high energies via primary and secondary
Fermi-like mechanisms~\citep{Pecora2018, Comisso2018}. In fully developed
GRMHD turbulence, accelerating islands and current sheets are present on
all scales and these could therefore provide the natural site for the
accelerating mechanism.

\begin{figure}
  \includegraphics[height=65mm,width=85mm]{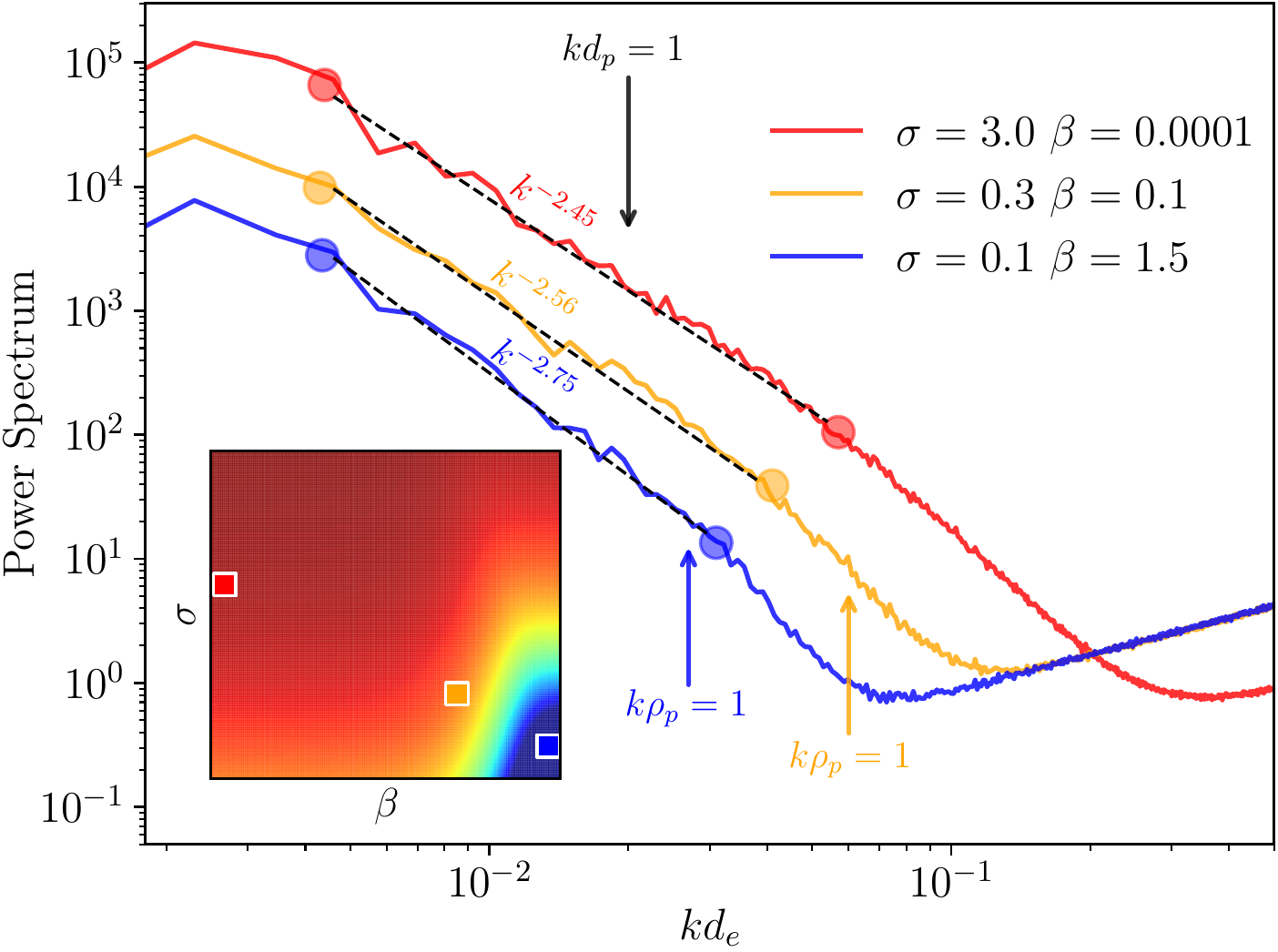}
  \caption{Magnetic-field power spectra for three simulations sampling
    important locations in the $(\beta, \sigma)$ space of
    parameters. Each simulation is marked with a different color and the
    corresponding location is shown in the inset, which reports also the
    electron spectral index. Black dashed lines indicate the turbulent
    power laws, while the circles delimit the boundaries of each
    turbulent range, which we define as the limits of the power-law
    scaling; the arrows mark the wavevectors associated to the
    proton-skin depth ($k d_p=1$) and to the proton Larmor radius
      ($k \rho_p = 1$), which is outside the horizontal scale for the red
      line.}
  \label{fig:5}
\end{figure}

In this simple picture, it is natural to expect that the larger the
spectrum of fluctuations at small scales, the more efficient the
accelerating mechanism~\citep{Haynes2014}. To validate whether this
applies also to trans-relativistic plasmas, we have computed the {(not
  normalized)} isotropic power spectrum of the magnetic field for three
representative simulations and reported them in Fig.~\ref{fig:5} as a
function of the dimensionless $k d_e$ [the inset shows with colored
  squares the location in the $(\sigma$, $\beta)$ plane of the three
  configurations, while the arrows mark the wavevectors associated
    to the proton-skin depth ($k d_p=1$) and to the proton Larmor radius
    ($k \rho_p = 1$)] and over a downsampled grid of $(1024)^2$ {(see
  Appendix for a discussion)}. In essence, after assuming the turbulence
to be isotropic and homogeneous, we integrate the 2D Fourier transforms
$\widetilde{B}_i$ over concentric shells (in this sense, the power
spectrum is isotropic) to obtain one-dimensional spectra, whose sum we
plot in Fig.~\ref{fig:5} [note that the growth of the power spectrum at
  large wavenumbers is a typical noise effect of PIC simulations due to a
  finite number of particles, \citep[see, \eg][]{Karimabadi2013}].

In general, Fig.~\ref{fig:5} reveals a number of interesting features,
moving in the parameter space from (low-$\beta$, high-$\sigma$) to
(high-$\beta$, low-$\sigma$). First, the power spectrum is clearly higher
in the case of the low-$\beta$, high-$\sigma$ simulation, confirming a
more efficient cascade process~\citep{Franci2016}.  {Second, the
  spectrum is shallower in the sub-ion inertial range
  \citep{Sahraoui2009} indicating a more developed turbulence}. Finally,
and more interestingly, the turbulent cascades terminate at much smaller
scales for (low-$\beta$, high-$\sigma$) simulations, {suggesting the
  existence of thinner current sheets at subproton scales that accelerate
  particles more efficiently~\citep{Pecora2018}}.

\section{Discussion and conclusions}

With the goal of gaining a deeper understanding of the properties of
plasmas near astrophysical compact objects, we have employed the PIC
\texttt{Zeltron} code to carry out a large campaign of {two-dimensional}
simulations of special-relativistic, decaying plasma turbulence in the
trans-relativistic regime. Particularly important in our analysis is the
use of a physical mass ratio between electrons and protons and the
exploration of a wide range of values in the plasma-$\beta$ parameter
($\beta=10^{-4}-1.5$) and in the magnetization $\sigma$
($\sigma=0.1-3.0$). Having simulated such a large portion of the space of
parameters encountered in astrophysical plasmas has allowed us to derive
{analytical fitting functions} for the behaviour of a number of important
plasma quantities as a function of $\beta$ and $\sigma$. More
specifically, we have presented 2D fitting functions of the electron
spectral index $\kappa(\beta,\sigma)$, of the efficiency in generating
nonthermal particles $\mathcal{E}(\beta,\sigma)$, and of the ratio
between the electron and proton temperatures
$\mathcal{T}(\beta,\sigma)$. These expressions provide compact and
reasonably accurate descriptions of the behaviour of these microphysical
plasma properties and can be employed in a number of scenarios involving
compact objects and described by macrophysical plasma
characteristics. {Importantly, since they have been derived from
  first-principle calculations, they represent a considerable improvement
  over the rather crude and purely empirical expressions employed
  at the moment in GRMHD simulations.} Finally, we have confirmed the
suggestion that plasmas with low $\beta$ and large $\sigma$ naturally
lead to broad turbulent scenarios and are the most efficient in
extracting particles from the thermal population and accelerating them
\citep{Pecora2018, Comisso2018}.

As these simulations represent one of the most systematic PIC
explorations of trans-relativistic turbulence, can be employed in a wide
range of astrophysical systems, such as jets and accretion disks around
supermassive black holes, and, of course, their imaging~\citep[see,
  \eg][]{EHT_M87_PaperV, EHT_SgrA_PaperV}. The formulas provided in
  this work can be improved by extending the present two-dimensional
  treatment to three dimensions and thus assessing the role played by
  dimensionality in studies of this type.

\begin{acknowledgments}
 
We thank the Referee for the useful comments that have improved our
presentation. This research is supported by the ERC Advanced Grant
``JETSET: Launching, propagation and emission of relativistic jets from
binary mergers and across mass scales'' (Grant No. 884631), by the
Deutsche Forschungsgemeinschaft (DFG, German Research Foundation) through
the CRC-TR 211 ``Strong-interaction matter under extreme conditions''
(project number 315477589), and by the State of Hesse within the Research
Cluster ELEMENTS (Project ID 500/10.006). LR acknowledges the Walter
Greiner Gesellschaft zur F\"orderung der physikalischen
Grundlagenforschung e.V. through the Carl W. Fueck Laureatus Chair. The
simulations were performed on HPE Apollo HAWK at the High Performance
Computing Center Stuttgart (HLRS) under the grant BNSMIC.

\end{acknowledgments}

{\appendix}

\begin{table*}
  \tiny
  \begin{center}
  \setlength{\tabcolsep}{1.0pt}
  \renewcommand{\arraystretch}{1.5}
  \begin{tabular}{l|rrrrrrrrrrrrrrrrrrrrrrrrrrrrrrrrrrrrrr}
    \hline
    \hline
    {Run} & 1 & 2 & 3 & 4 & 5 & 6 & 7 & 8 & 9 & 10 & 11 & 12 & 13 & 14 & 15 & 16 & 17 & 18 & 19 & 20 & 21 & 22 & 23 & 24 & 25 & 26 & 27 & 28 & 29 & 30 & 31 & 32 & 33 & 34 & 35\\
    \hline {$\sigma$} & 1.0{e}-1 & 1.0{e}-1 & 1.0{e}-1 & 1.0{e}-1 & 1.0{e}-1 & 1.0{e}-1 & 1.0{e}-1 & 1.0{e}-1 & 1.0{e}-1 & 1.0{e}-1 & 1.0{e}-1 & 3.0{e}-1 & 3.0{e}-1 & 3.0{e}-1 & 3.0{e}-1 & 3.0{e}-1 & 3.0{e}-1 & 3.0{e}-1 & 3.0{e}-1 & 3.0{e}-1 & 1.0{e}0 & 1.0{e}0 & 1.0{e}0 & 1.0{e}0 & 1.0{e}0 & 1.0{e}0 & 1.0{e}0 & 1.0{e}0 & 3.0{e}0 & 3.0{e}0 & 3.0{e}0 & 3.0{e}0 & 3.0{e}0 & 3.0{e}0 & 3.0{e}0 \\
    \hline {$\beta$} & 1.0{e}-4 & 3.0{e}-4 & 1.0{e}-3 & 3.0{e}-3 & 1.0{e}-2 & 2.0{e}-2 & 1.0{e}-1 & 3.0{e}-1& 7.0{e}-1 & 1.0{e}0 & 1.5{e}0 & 1.0{e}-4 & 3.0{e}-4 & 1.0{e}-3 & 3.0{e}-3 & 1.0{e}-2 & 3.0{e}-2 & 1.1{e}-1 & 3.4{e}-1 & 5.5{e}-1 & 1.0{e}-4 & 3.0{e}-4 & 1.0{e}-3 & 3.0{e}-3 & 1.0{e}-2 & 3.0{e}-2 & 1.0{e}-1 & 1.6{e}-1 & 1.0{e}-4 & 3.0{e}-4 & 1.0{e}-3 & 3.0{e}-3 & 1.0{e}-2 & 2.6{e}-2 & 5.5{e}-2\\
    \hline {$\theta_p$} & 5.0{e}-6 & 1.5{e}-5 & 5.0{e}-5 & 1.5{e}-4 & 5.0{e}-4 & 1.0{e}-3 & 5.0{e}-3 & 2.0{e}-2 & 4.5{e}-2 & 6.8{e}-2 & 1.0{e}-1 & 1.5{e}-5 & 5.0{e}-5 & 1.5{e}-4 & 5.0{e}-4 & 1.5{e}-3 & 5.0{e}-3 & 2.0{e}-2 & 8.0{e}-2 & 2.0{e}-1 & 5.0{e}-5 & 1.5{e}-4 & 5.0{e}-4 & 1.5{e}-3 & 5.0{e}-3 & 1.5{e}-2 & 5.0{e}-2 & 2.0{e}-1 & 1.5{e}-4 & 5.0{e}-4 & 1.5{e}-3 & 5.0{e}-3 & 1.5{e}-2 & 5.0{e}-2 & 2.0{e}-1\\
    \hline {$\theta_e$} & 9.2{e}-3 & 2.7{e}-2 & 9.2{e}-2 & 2.7{e}-1 & 9.2{e}-1 & 1.8{e}0 & 9.2{e}0 & 3.7{e}1 & 8.3{e}1 & 1.2{e}2 & 1.8{e}2 & 2.7{e}-2 & 9.2{e}-2 & 2.7{e}-1 & 9.2{e}-1 & 2.7{e}0 & 9.2{e}0 & 3.7{e}1 & 1.5{e}2 & 3.7{e}2 & 9.2{e}-2 & 2.7{e}-1 & 9.2{e}-1 & 2.7{e}0 & 9.2{e}0 & 2.7{e}1 & 9.2{e}1 & 3.7{e}2 & 2.7{e}-1 & 9.2{e}-1 & 2.7{e}0 & 9.2{e}0 & 2.7{e}1 & 9.2{e}1 & 3.7{e}2\\ 
	\hline {$\lambda_{_D}$} & 9.6{e}-2 & 1.6{e}-1 & 3.0{e}-1 & 5.2{e}-1 & 9.6{e}-1 &
	1.3{e}0 & 3.0{e}0 & 6.1{e}0 & 9.2{e}0 & 1.1{e}1 & 1.3{e}1 & 1.6{e}-1 &
	3.0{e}-1 & 5.2{e}-1 & 9.6{e}-1 & 1.6{e}0 & 3.0{e}0 & 6.1{e}0 & 1.2{e}1&
	1.9{e}1 & 3.0{e}-1 & 5.2{e}-1 & 9.6{e}-1 & 1.6{e}0 & 3.0{e}0 & 5.2{e}0 &
	9.6{e}0 & 1.9{e}1 & 5.2{e}-1 & 9.6{e}-1 & 1.6{e}0 & 3.0{e}0 & 5.2{e}0 &
	9.6{e}0 & 1.9{e}1 \\ \hline
  \end{tabular}
  \caption{Summary of the physical parameters of our main simulations,
    which are all performed with the real electron-to-proton mass ratio,
    equal electron and proton initial temperatures, a resolution of three
    cells per electron-skin depth ($d_e/dx=3$), and a box of size $\sim
    5461\,d_e$ in both directions. From top to bottom we report: the
    number of the Run, the magnetization $\sigma$, the plasma $\beta$,
    the dimensionless temperatures $\theta_{p,e}$ for protons and
    electrons respectively, and the Debye length $\lambda_D$ in
      units of $d_e$. In all our simulations we have initialized each
    computational cell with 10 particles (5 protons and 5 electrons).}
    \label{table1}
  \end{center}
\end{table*}

\begin{table*}
    \setlength{\tabcolsep}{2.0pt}
    \renewcommand{\arraystretch}{1.5}
    \begin{tabular}{l|rrrrrrrrrrrrrrrrrrrrrrrrrrrrrrrrrrrrrr}
      \hline
      \hline
	  {Run} & A1 & A2 & A3 & B1 & B2 & B3 \\
	  \hline {$\sigma$} & 3.0{e}-1 & 3.0{e}-1 & 3.0{e}-1 & 3.0{e}-1 & 3.0{e}-1 & 3.0{e}-1 \\
	  \hline {$\beta_p$} & 3.0{e}-4 & 3.0{e}-4 & 3.0{e}-4 & 3.0{e}-4 & 3.0{e}-4 & 3.0{e}-4 \\
	  \hline {$\beta_e$} & 3.0{e}-2 & 3.0{e}-3 & 3.0{e}-5 & 3.0{e}-4 & 3.0{e}-4 & 3.0{e}-4  \\
	  \hline {$\theta_p$} & 5.0{e}-5 & 5.0{e}-5 & 5.0{e}-5 & 5.0{e}-5 & 5.0{e}-5 & 5.0{e}-5  \\ 
	  \hline {$\theta_e$} & 9.18{e}0 & 9.18{e}-1 & 9.18{e}-3 & 9.18{e}-2 & 9.18{e}-2 & 9.18{e}-2  \\
	  \hline {$\mathcal{T}_0$} & 1.0{e}-2 & 1.0{e}-1 & 1.0{e}+1 & 1.0{e}0 & 1.0{e}0 & 1.0{e}0 \\
	  \hline {$d_e/dx$} & 3.0{e}0 & 3.0{e}0 & 3.0{e}0 & 3.0{e}0 & 6.0{e}0 & 1.2{e}+1  \\ 
	  \hline {$L/d_e$} & 5.46{e}+3 & 5.46{e}+3 & 5.46{e}+3 & 2.73{e}+3
          & 2.73{e}+3 & 2.73{e}+3 \\
             \hline
    \end{tabular}
    \caption{Table of simulation in which we varied different
      parameters. Runs A1-A3 have different initial $\mathcal{T} =
      T_p/T_e$ (and hence different $\beta_{e}$ and $\theta_e$), while
      all other parameters ($\sigma, \beta_p, \theta_p, d_e/dx, L/d_e$)
      are the same. Runs B1-B3 have different values of the electron-skin
      depth per $dx$ and use a smaller physical box of $2730\,d_e$. From
      top to bottom we report: the number of the Run, the magnetization
      $\sigma$, the proton and electron plasma $\beta$, the proton and
      electron dimensionless temperatures $\theta_{p,e}$, the initial
      temperature ratio $\mathcal{T}_0$, the number of cells per
      electron-skin depth ($d_e/dx$), and the physical box size in terms
      of electron-skin depth.}
    \label{table2}
\end{table*}

In what follows, we provide additional information on our analysis
concentrating on three specific aspects: a detailed summary of the
properties of the simulations carried out in the campaign, the evidence
that stationarity is reached when extracting the spectral information,
and a comparison of simulations with different resolutions.

{\section*{Summary of simulations}}

Our systematic investigation of the $\beta,\sigma$ space of parameters
consists of 35 large-scale, high-resolution simulations whose main
properties are reported in Table \ref{table1}. All these simulations were
performed in two spatial dimensions with the real electron-to-proton mass
ratio, a physical-box size of $L \sim 5461\,d_e$ (where $d_e$, we recall,
is the electron-skin depth) in each of the two spatial directions, and
the same electron-to-proton initial temperature, \ie $\mathcal{T}_0=1$.
In addition, we have performed six simulations with varying properties
with respect to the main ones and reported in Table \ref{table2}.

As a first test, to show that our final configuration is independent of
the initial electron-to-proton temperature, we have varied
$\mathcal{T}_0$ spanning in the range $[0.001-10.0]$ (see Runs A1-A3 in
Table \ref{table2}, see Fig.\ref{fig:1}). Note that for these
configurations, the plasma $\beta$ is different for electrons and
protons. Next, we checked that our results are insensitive to the choice
of different (higher) resolutions in terms of $d_e/dx$, increasing the
resolution up to ${d_e/dx = 12}$ (see Runs B1-B3 in
Table \ref{table2}). In the latter case, we have used a physical box of
$L/d_e = 2730$ in both directions and varied the number of mesh points
from $(8192)^2$ up to $(32768)^2$. In this last high-resolution
configuration, we have followed the dynamics of $\sim 1.1 \times 10^{11}$
particles.

{\section*{Stationarity of spectra}}

\begin{figure}
  \centering
  \includegraphics[width=0.8\columnwidth]{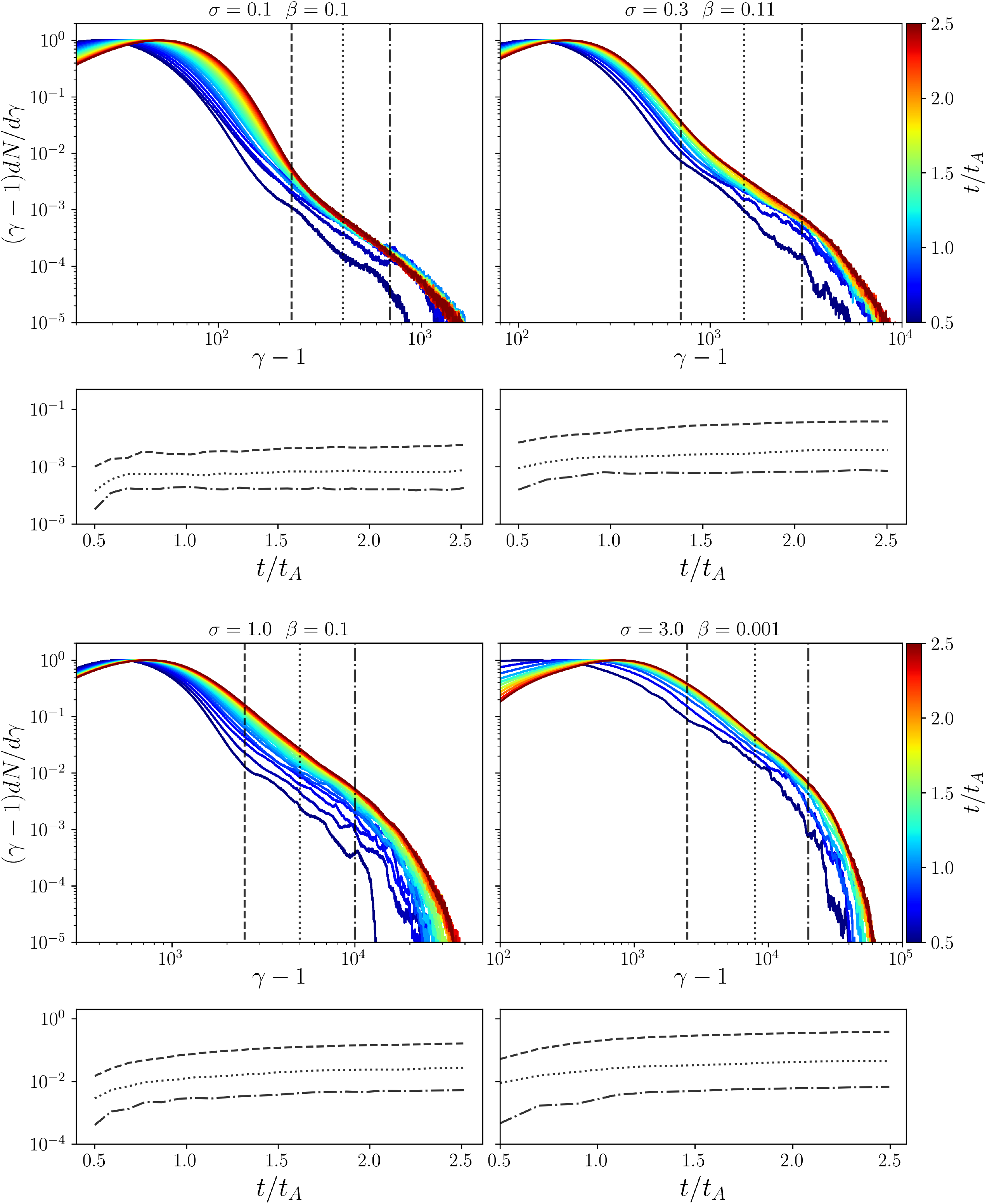}
  \caption{Four representative simulations in which we show the
    stationarity of the electron-energy spectra (see Runs 7, 18, 27, and
    31 in Table \ref{table2}). For each simulation, we report the spectra
    at different times during the evolution as indicated by the colormap
    on the right of each of the four panels. Marked with black vertical
    lines of various type are three different values of the Lorentz
    factor $\gamma-1$ and the corresponding evolutions are shown in the
    bottom panels for each of the four simulations considered. Clearly,
    all cases show that by $t/t_A \sim 2.0$ the simulations have reached
    stationarity.}
  \label{fig:1a}
\end{figure}

Next, we provide evidence that the computed electron-energy spectra reach
a steady state after $t/t_A \gtrsim 1.8-2.0$, so that the extraction of
the spectral index $\kappa$ and of the efficiency $\mathcal{E}$ is both
accurate and robust. Figure~\ref{fig:1a} shows four representative
simulations having different values of $\sigma$ (see Runs 7, 18, 27, and
31 in Table \ref{table1}). In each case, we plot the electron-energy
spectra at different times during the evolution as indicated by the
colormap on the right of each of the four panels. Furthermore, marked
with black vertical lines of various type are three different values of
the Lorentz factor $\gamma-1$ and the corresponding evolutions are shown
in the bottom panels for each of the four simulations
considered. Clearly, all cases show that by $t/t_A \sim 2.0$ the
simulations have reached stationarity with relative time variations that
are $\lesssim 1.5\%$, so that $\kappa$ and $\mathcal{E}$ can be extracted
reliably.

{\section*{Resolution Tests}}

\begin{figure}
  \centering
  \includegraphics[width=0.6\columnwidth]{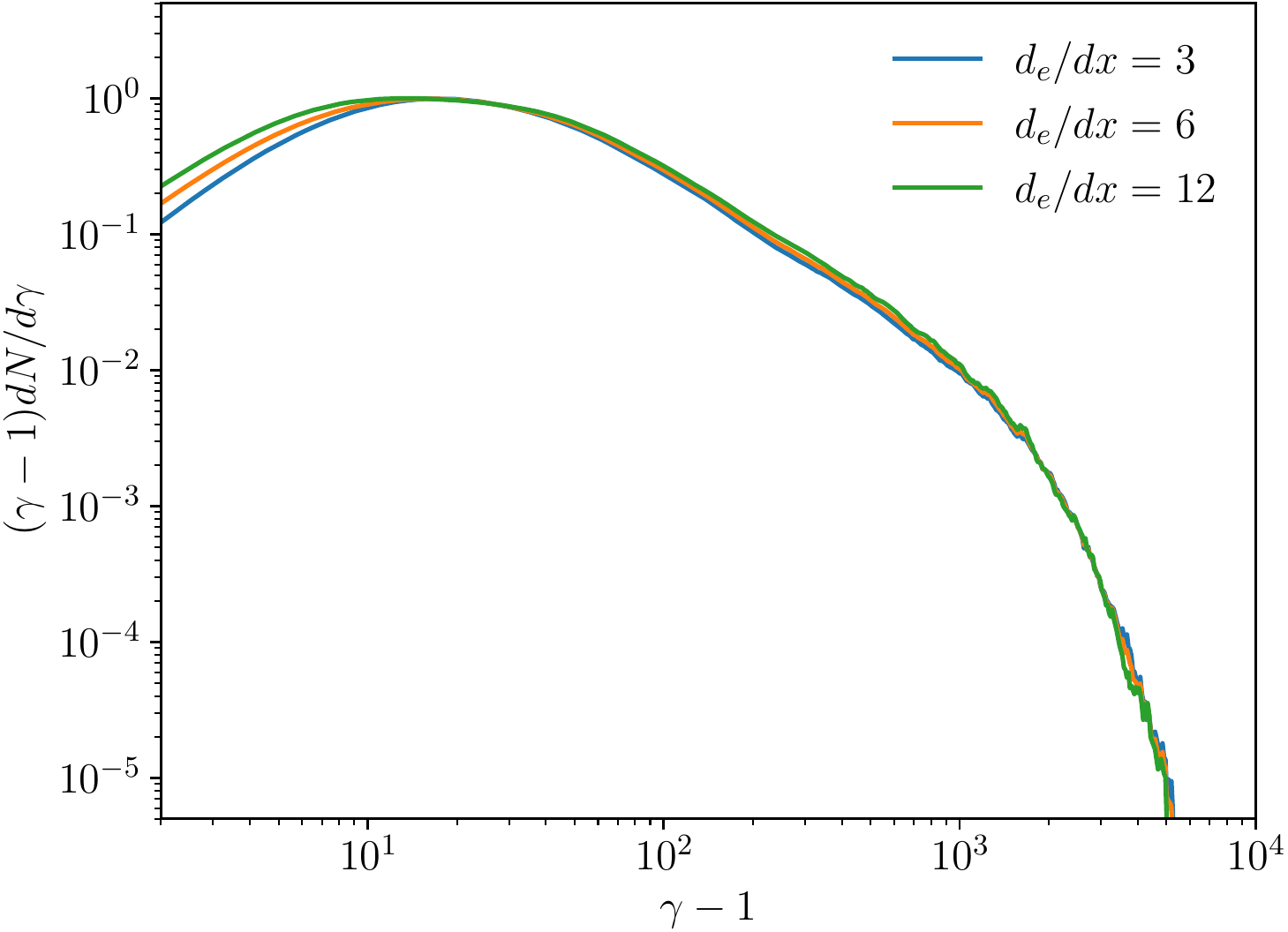}
  \caption{Electron-energy spectra with $\sigma=0.3$ and $\beta=3 \times
    10^{-4}$ for three different resolutions $d_e/dx=3, 6$ and $12$,
    using a physical box size of $L/d_e = 2730$. The spectra are computed
    at $t/t_A=2.0$ and clearly show to be nearly insensitive to
    the increased resolution.}
  \label{fig:2a}
\end{figure}

Finally, we have verified that our results are insensitive to the choice
of spatial resolution. In particular, we have performed three simulations
using an increasing number of cells per electron skin depth, from
$d_e/dx=3$ up to $d_e/dx=12$ (see Runs B1-B3 in Table \ref{table2}).
Figure~\ref{fig:2a} compares the electron-energy spectra for a case with
$\sigma=0.3$ and $\beta=3 \times 10^{-4}$ when varying the number of
electron-skin depths per cell, \ie $d_e/dx=3-12$. Clearly, the main
features of the electron-energy spectra and in particular the slope are
very similar for the three different resolutions. Indeed, the relative
differences between the three spectra are $\lesssim 6.0\%$ and thus even
smaller than the variations due to the stochastic nature of turbulence,
which can cause variation in $\kappa$ up to $\sim 10.0 \%$
\citep{Ball2018a}.

\begin{figure}
	\centering
	\includegraphics[width=0.6\columnwidth]{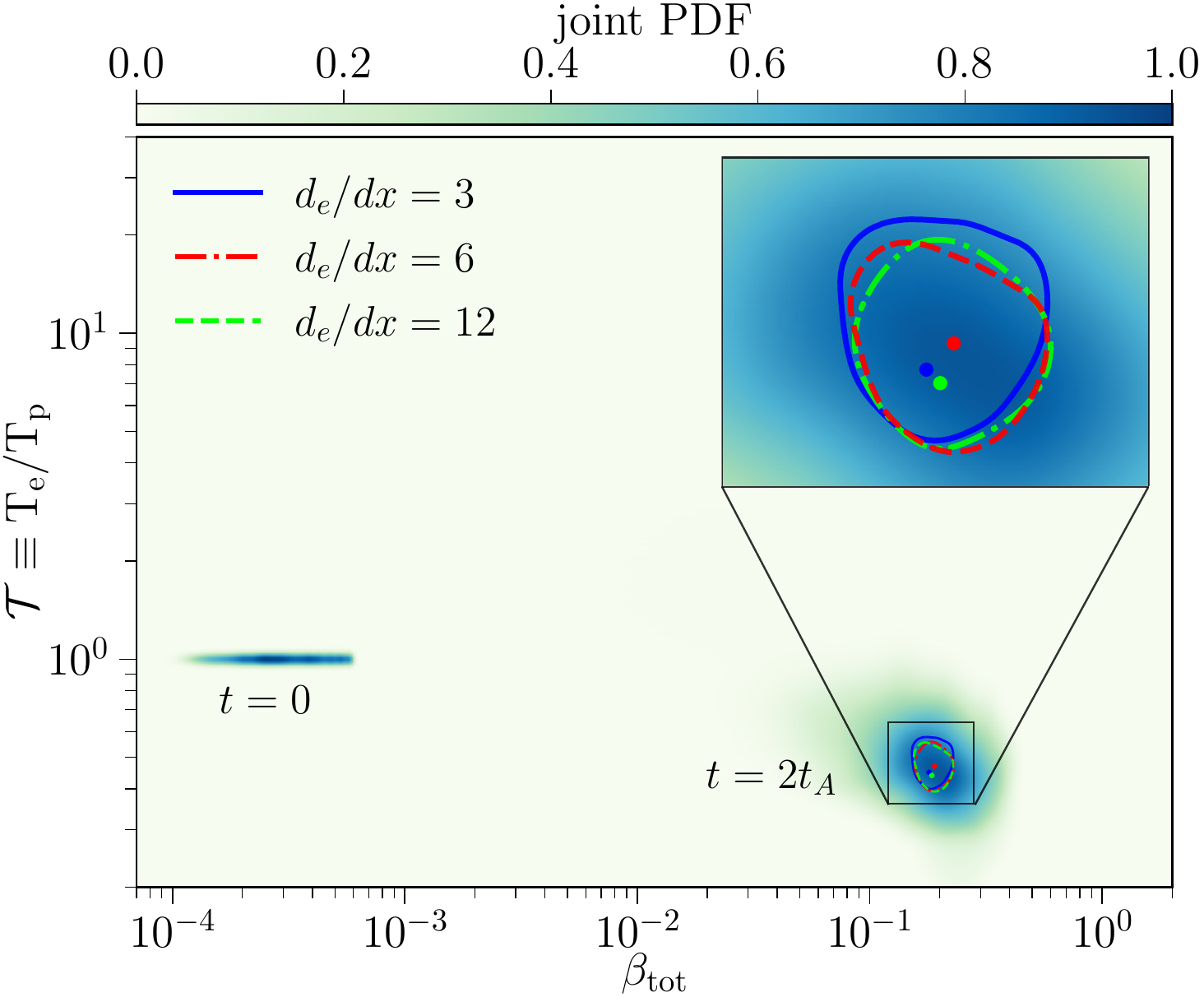}
	\caption{Initial and final values at time $t=2 \, t_A$ of the normalized joint PDFs of the temperature ratio $\mathcal{T}$ and of the total plasma $\beta$ (see Figure 1), using three different resolutions, namely $d_e/dx = 3, 6$, and $12$. The inset shows the $90 \%$ contour lines of the joint PDFs, while the circle mark the maxima of each distribution. Note that all PDFs converge to the same final area in the $(\mathcal{T},\beta_{\text{tot}})$ plane.} 
	\label{fig:3a}
\end{figure}

In Figure \ref{fig:3a} we show the joint PDFs for the ratio of temperatures $\mathcal{T}$ and the plasma $\beta_{\text{tot}} = \beta_e + \beta_p$ for the same runs. In the inset we report a zoom-in of the central region of the PDFs at the final time of $t=2\, t_A$. The color-coded contour report the $90 \%$-value for each distribution, while the circle represent the maximum of each joint PDF. One can see that for the three different resolutions we obtain similar final distributions, with a variation in $\mathcal{T} \lesssim 5.0 \%$.

\medskip

As a concluding remark, we note that the power spectrum in
Fig.~\ref{fig:5} has been computed on a down-sampled grid of
$(1024)^2$ points and not on the full-resolution data of $(16348)^2$
points. This coarse-graining operation is routinely done in such
expensive simulations, and for two distinct
reasons. First, the large particle noise due to the high temperatures
reached essentially blurs out the smallest scales, so that using the full
resolution does not really provide any additional information. Second,
the downsampling allows us to reduce by a factor of $16^2 \sim 250$ the
space needed for the ouput (we recall that we save data for 38 fields at
very high cadence). As a result, while the simulation maximum wavenumber
is $k_{\rm max}\,d_e = 9.4$ and is not shown in the spectrum in
Fig.~\ref{fig:5}, the maximum wavenumber in the downsampled spectrum is
$k_{\rm max}\,d_e = 0.6$ and is well-captured.

\end{document}